\documentclass[fleqn,twoside]{article}
\usepackage{espcrc2}
\usepackage{amsmath,amssymb}

\usepackage{graphicx}
\newcommand{\vecr}{{\bf r}}

\title{Oscillator strength distribution in C$_3$H$_6$ isomers studied with
the time-dependent density functional method in the continuum}

\author{Takashi Nakatsukasa\address[Tohoku]
        {Physics Department, Tohoku University, Sendai 980-8578, Japan}
        \thanks{Email: takashi@nucl.phys.tohoku.ac.jp}
        and
        Kazuhiro Yabana\address[Tsukuba]
        {Institute of Physics, University of Tsukuba, Tsukuba 305-8571, Japan}
        \thanks{Email: yabana@nucl.ph.tsukuba.ac.jp}
       }
       
\begin{document}

\begin{abstract}
We present photoabsorption oscillator strengths for C$_3$H$_6$ molecules
with emphasis on the difference between isomers, cyclopropane and propylene.
We use an iterative numerical method based on the time-dependent local density
approximation with continuum, which we have recently developed.
The oscillator strengths for the two isomers differ at photon energies above
their ionization thresholds.
The magnitude and the shape of the oscillator strength distribution are in
good agreement with recent experiments.
The differences between the isomers arise from difference in symmetry
of electronic states and different behaviors of continuum excitations.
\vspace{1pc}
\end{abstract}

\maketitle

\section{Introduction}
\label{sec:intro}

The photoabsorption and photoionization cross sections of molecules
are of significant interest in many fields of both fundamental
and applied sciences.
The oscillator strength distribution characterizing the optical response
is the most important quantity in understanding the interaction of photon with
electrons in atoms, molecules, and matters.
The oscillator strength distribution in the whole spectral region
has been extensively studied with
the advanced synchrotron radiation and the high resolution
electron energy loss spectroscopy \cite{Ber79,Hat99}.
In order to see how the oscillator strength changes with varying
molecular structures, it is useful to study isomer molecules.
Since the isomers consist of the same kind and the same number of atoms,
we expect similarity of the oscillator strengths at high photon energies.
This is because the molecular structure has little influence on the
excitation of inner core electrons.
However, the valence photoabsorption may differ according to the
difference of electronic states between the isomers.
In fact, Koizumi et al. have observed a prominent distinction for
the cross sections of simple hydrocarbon isomers, C$_3$H$_6$
(cyclopropane and propylene), at photon energies of $10-20$ eV
\cite{Koi85}.
The photoabsorption and photoionization data of cyclopropane in this energy
region were later improved by a measurement with metallic thin film
windows \cite{Kam99}.
These works clearly show that the oscillator strengths have different
peaks and shoulders depending on the isomers in a continuous spectral
region above the ionization potentials (IPs).

Theoretical investigation for the isomer effect of C$_3$H$_6$ has been
demanded for a long time, however, none has been reported so far.
This is due to difficulties in treatment of the electronic continuum
in a non-spherical multicenter potential.
There are several methods which are able to take into
account correlations among valence electrons in the continuum
\cite{LS84,CCRM91,WL99,Lan80,YM85}.
Nevertheless,
some of them are not suitable for calculating detailed structures of
the spectra, and some are difficult to be applied to large molecules.
We have recently developed an alternative theoretical method for this
purpose \cite{NY01}.
The method is based on the time-dependent local-density approximation (TDLDA)
in a grid representation of the three-dimensional Cartesian coordinate,
and utilizes the Green's function to take account of the continuum
boundary condition.
We use iterative methods to solve linear algebraic equations
for construction of the dynamical screening field above the first IP.
The theoretical background of our method is similar to the one of
Ref.~\cite{LS84} in which the authors used a single-center expansion technique.
However, the application was limited to small axially symmetric molecules,
because of difficulties in the single-center expansion.
Our method is based on direct calculation of the self-consistent screening
potential in the three-dimensional grid
representation, which does not rely on the expansion
and requires no spatial symmetry.

In the present Letter, we report the valence photoabsorption of the
C$_3$H$_6$ isomers studied with the continuum TDLDA method in Ref.~\cite{NY01},
and would like to show the power of the method.
Then, we give an interpretation of the continuous spectra and
elucidate origins of the isomer effects.

\section{Theory and computational method}
\label{sec:method}

Optical response of molecules is characterized by the oscillator strength,
denoted as $df/d\omega$ in the followings, which is given by
\begin{equation}
\label{dfdw}
\frac{df}{d\omega}= -\frac{2m\omega}{3\pi}{\rm Im}
        \sum_{\nu=x,y,z} \int d^3r r_\nu \delta n_\nu({\vecr},\omega) ,
\end{equation}
where the transition density $\delta n_\nu$ is related to
the Fourier component of a time-dependent external dipole perturbation in
$\nu$-direction, $V_\nu(\omega)=r_\nu$,
through a complex susceptibility
($\delta n_\nu(\omega) = \chi(\omega) V_\nu(\omega)$).

The TDLDA describes a spin-independent $N$-electron system in terms of
the time-dependent Kohn-Sham equations.
Correlations among electrons are taken into account through
deformations of the self-consistent Kohn-Sham potential.
Linearizing the Kohn-Sham potential with respect to the transition density,
we obtain
\begin{equation}
\label{delta_n}
\begin{split}
\delta n_\nu(\vecr,\omega) &= \int d^3 r' \chi_0(\vecr,\vecr';\omega)\\
    & \left\{ r'_\nu + \int d^3r''
     \frac{\delta V_{\rm KS}[n(\vecr')]}{\delta n(\vecr'')}
      \delta n_\nu(\vecr'') \right\}.
\end{split}
\end{equation}
The $\chi_0(\vecr,\vecr';\omega)$ is a complex susceptibility for a system
without the correlations and is given by
\begin{equation}
\label{chi_0}
\begin{split}
\chi_0(\vecr,\vecr';\omega) =
 2\sum_i^{\rm occ} &
 \phi_i(\vecr) \left\{
     \left( G(\vecr,\vecr';\epsilon_i-\omega^*)\right)^* \right. \\
   &+ \left. G(\vecr,\vecr';\epsilon_i+\omega) \right\}
   \phi_i(\vecr') .
\end{split}
\end{equation}
Here, $\phi_i$'s are the ground-state Kohn-Sham orbitals and  
$G$ is the Green's function for an electron in the static
Kohn-Sham potential.
In order to properly treat the electronic continuum,
the outgoing boundary condition must be imposed on $G$.
Construction of the Green's function is easily done for
a rotationally invariant potential $V_0(r)$,
using the partial wave expansion \cite{ZS80}.
Thus, we split the Kohn-Sham potential into two parts,
a long-range spherical part, $V_0(r)$, and a short-range deformed part,
$\tilde{V}(\vecr)=V_{\rm KS}(\vecr)-V_0(r)$.
First, we construct the Green's function, $G_0$, for the spherical
potential $V_0$, then, $G$ can be obtained from an identity
\begin{equation}
\label{G}
G = G_0 + G_0 \tilde{V} G .
\end{equation}

We solve Eqs.~(\ref{delta_n}), (\ref{chi_0}), and (\ref{G}) simultaneously
in the uniform grid representation of the three-dimensional real space.
Equations (\ref{delta_n}) and (\ref{G}) are linear algebraic equations
with respect to $\delta n$ and $G$, respectively,
for which an iterative method provides an efficient algorithm for
the numerical procedure.
We adopt the generalized conjugate residual method for these non-hermitian
problems.
We would like to refer the reader to Ref.~\cite{NY01}
for detailed discussion of
the methodology and the theoretical background.

The exchange-correlation potential is a sum of the local density part
given by Ref.~\cite{PZ81}, $\mu^{\rm (PZ)}[\rho]$,
and the gradient correction of Ref.~\cite{LB94}, 
$\mu^{\rm (LB)}[\rho,\nabla\rho]$, which will be abbreviated to LB potential.
This gradient correction is constructed so as to reproduce the correct 
Coulomb asymptotic behavior of the potential ($-e^2/r$) and to describe
the Rydberg states.
It was also pointed out that the LB potential is 
necessary to reproduce the excitation energies of high-lying bound states
in the TDLDA \cite{Cas98}.
In our previous work \cite{NY01},
we have also found for simple molecules that the TDLDA with the LB potential
reasonably accounts for resonances embedded in the continuum.

%%%%%%%%%%%%%%%%%%%%%%%%%%%%%%%%%%%%%%%%%%%%%%%%%%%%%%%%%%%%%%%%%%
\begin{table*}
\caption{Calculated eigenvalues of occupied valence orbitals in units of eV.
Values in brackets indicate eigenvalues calculated
using a different value of the parameter in $\mu^{\rm (LB)}$ ($\beta=0.05$).
See text for details.
Line styles in Figs.~\ref{dfdw_cyclo1} (b), \ref{dfdw_propy} (b), and
\ref{dfdw_cyclo2} (b) are indicated in the third and sixth columns.
We use following abbreviations: ``S'' for the solid, ``Do'' for the dotted,
``Da'' for the dashed, ``LD'' for the long-dashed, ``DD'' for the dot-dashed,
and ``T-'' for the thick lines.
}
\begin{center}
\begin{tabular}{crc|crc}
\hline
\multicolumn{3}{c|}{Cyclopropane} & \multicolumn{3}{c}{Propylene}\\
Orbital      & \multicolumn{1}{c}{Calc.}   & Line & Orbital    & Calc.  & Line\\
\hline
$(3e')^4$    & $-10.6$ $(-11.9)$ & T-S &
 $(2a'')^2$ & $ -9.9$ & T-S \\
$(1e'')^4$   & $-11.9$ $(-13.2)$ & Do  &
 $(10a')^2$ & $-11.4$ & T-Do \\
$(3a_1')^2$  & $-14.8$ $(-16.1)$ & S   &
 $(9a')^2$  & $-12.0$ & T-Da \\
$(1a_2'')^2$ & $-15.4$ $(-16.8)$ & Da  &
 $(1a'')^2$ & $-13.4$ & T-LD \\
$(2e')^4$    & $-17.5$ $(-19.0)$ & DD  &
 $(8a')^2$  & $-13.6$ & DD \\
$(2a_1')^2$  & $-23.9$ $(-25.3)$ & LD  &
 $(7a')^2$  & $-14.6$ & S \\
             &                   &     &
 $(6a')^2$  & $-16.6$ & Do \\
             &                   &     &
 $(5a')^2$  & $-19.6$ & Da \\
             &                   &     &
 $(4a')^2$  & $-22.3$ & LD \\
\hline
\end{tabular}
\end{center}
\label{spe}
\end{table*}
%%%%%%%%%%%%%%%%%%%%%%%%%%%%%%%%%%%%%%%%%%%%%%%%%%%%%%%%%%%%%%%%%%
\section{Results and discussion}
\label{sec:results}

\subsection{Ground-state properties}
We fix the geometry of nuclei optimized for the ground state.
This is based on a semiempirical
method known as PM3 \cite{Ste89}.
%The nuclear coordinates are given in Table~\ref{geometry}.
We only treat valence electrons in the TDLDA calculation.
Thus, we use the norm-conserving pseudopotential \cite{TM91}
with a separable approximation \cite{KB82} for the electron-ion potentials.
The coordinate space is discretized in a square mesh of 0.3 \AA\ and we adopt
all the grid points inside a sphere of 6 \AA\ radius.
This results in a model space of 33,401 grid points.

First, we calculate the ground state of cyclopropane and propylene
by solving the Kohn-Sham equations with the exchange-correlation potential of
$\mu^{\rm (PZ)}+\mu^{\rm (LB)}$.
The LB potential $\mu^{\rm (LB)}$ contains a parameter $\beta$ \cite{LB94},
and we adjust this value to make eigenvalues of the highest occupied molecular
orbitals (HOMO) coincide with the empirical vertical IPs
($10.54$ eV for cyclopropane \cite{Ple81}
and $9.91$ eV for propylene \cite{Kra78}).
The occupied valence orbitals in the ground state calculated with
$\beta=0.015$ are listed in Table~\ref{spe}.
The HOMO eigenvalues are well reproduced for both isomers.
Propylene has a geometry of the $C_s$ point group while
cyclopropane has the $D_{3h}$ group.
Although these isomers possess equal number of valence electrons
(eighteen valence electrons),
the electron configuration of cyclopropane is
more degenerate in energy because of the higher symmetry.

\subsection{Photoabsorption oscillator strength}
\begin{figure}
\centerline{\includegraphics[width=0.4\textwidth]{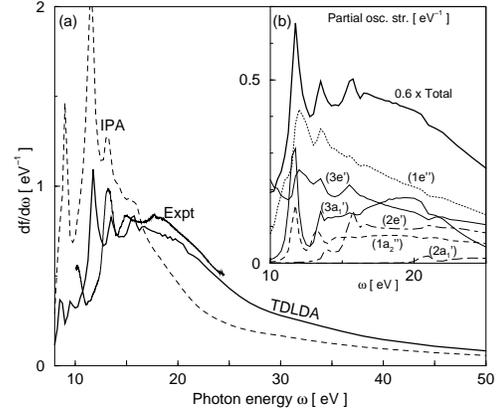}}
\caption{(a) Calculated (thick solid line) and experimental (thin solid)
photoabsorption oscillator strength distribution
for a cyclopropane molecule as a function of photon energy.
The dashed line indicates the IPA calculation without dynamical
screening effects.
The experimental data are taken from Ref.~\cite{Kam99}.
See text for details.
(b) An energy region of $10<\omega<25$ eV is magnified and the total
oscillator strength is decomposed into those associated with
different occupied valence electrons.
See Table~\ref{spe} for correspondence between a line style and
an occupied orbital.
}
\label{dfdw_cyclo1}
\end{figure}

\begin{figure}
\centerline{\includegraphics[width=0.4\textwidth]{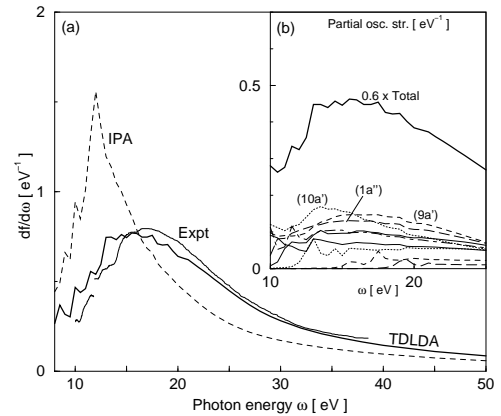}}
\caption{The same as Fig.~\ref{dfdw_cyclo1} but for propylene.
The experimental data are taken from Ref.~\cite{Koi85}.
}
\label{dfdw_propy}
\end{figure}

Now we calculate the photoresponse of the isomers.
We use complex frequencies, $\omega+i\Gamma/2$ with $\Gamma=0.5$ eV.
The $\Gamma$ plays a role of a smoothing parameter
to make the energy resolution finite.
This also helps a convergence of the numerical iteration procedure \cite{NY01}.
In Figs.~\ref{dfdw_cyclo1} and \ref{dfdw_propy} respectively,
we show the calculated photoabsorption
oscillator strength, $df/d\omega$, for cyclopropane and propylene
in a frequency (photon energy) range of $8-50$ eV.
The calculations have been done with a frequency mesh of
$\Delta\omega=0.25$ eV for a region of $8\leq\omega\leq 20$ eV,
and with $\Delta\omega=0.5$ eV for the rest
of frequencies.
The oscillator strength distributions of the isomers are
nearly identical at $\omega \gtrsim 22$ eV.
This energy roughly corresponds to the ionization energy of the
lowest-lying $\sigma$ orbital.
The $df/d\omega$ monotonically decreases as the frequency increases
but has a large tail at high frequency.
This behavior of the high-frequency tail in $df/d\omega$ is one of
the characteristics of the electronic
excitations in the continuum, which was also found in our previous
studies of simple molecules \cite{NY01}.
The molecular structure has a little influence on
the electronic continuum in this energy region ($\omega\gtrsim 22$ eV).

In contrast, in the frequency region below 22 eV,
different structures are observed among the isomers.
The $df/d\omega$ of propylene shows a single broad peak at
$\omega=13\sim18$ eV with small wiggles.
On the other hand, distinctive three peaks at $\omega=11.8$, 13.5, and
15.7 eV, are found in cyclopropane.
This difference exactly matches the experimental findings of the
isomer effect (the thin solid line in Fig.~\ref{dfdw_cyclo1} (a)).
The energy positions of calculated peaks are lower than the experimental
ones by about 1.5 eV.
This is also true for propylene in Fig.~\ref{dfdw_propy} (a),
in which the broad peak is shifted
to lower energy by 1.5 eV compared to the experiment.
We would like to mention that,
if we treat the electrons as responding independently to the external
dipole field, we cannot reproduce the main feature of the oscillator strength
distributions.
We call this approximation ``independent-particle approximation (IPA)'',
which corresponds to neglecting the induced screening
potential, the second term in the bracket in Eq.~(\ref{delta_n}).
In Figs.~\ref{dfdw_cyclo1} (a) and \ref{dfdw_propy} (a), the IPA calculations
are shown by dashed lines.

The Thomas-Kuhn-Reiche (TRK) sum rule tells us the integrated oscillator
strength $f(\infty)=18$ in our calculations,
since we only treat valence electrons in C$_3$H$_6$.
In Table~\ref{TRK}, partial sum values of the oscillator strengths are
listed and compared to the experiment.
Again, the IPA calculation cannot account for the data, while the TDLDA
in the continuum well agrees with the experiment.
Calculated total sum values for $8<\omega<60$ eV are 16.2 for cyclopropane
and 15.8 for propylene.
These values correspond to about 90 \% of the TRK sum rule for valence
electrons.
%%%%%%%%%%%%%%%%%%%%%%%%%%%%%%%%%%%%%%%%%%%%%%%%%%%%%%%%%%%%%%%%%%
%%%%%%%%%%%%%%%%%%%%%%%%%%%%%%%%%%%%%%%%%%%%%%%%%%%%%%%%%%%%%%%%%%

%%%%%%%%%%%%%%%%%%%%%%%%%%%%%%%%%%%%%%%%%%%%%%%%%%%%%%%%%%%%%%%%%%
%%%%%%%%%%%%%%%%%%%%%%%%%%%%%%%%%%%%%%%%%%%%%%%%%%%%%%%%%%%%%%%%%%

%%%%%%%%%%%%%%%%%%%%%%%%%%%%%%%%%%%%%%%%%%%%%%%%%%%%%%%%%%%%%%%%%%
\begin{table*}
\caption{Partially summed oscillator strengths for C$_3$H$_6$ isomers.
The experimental values are estimated from data in Refs.~\cite{Koi85,Kam99}.
}
\begin{center}
\begin{tabular}{l|rrrr}
\hline
      & \multicolumn{2}{|c}{Cyclopropane}  & \multicolumn{2}{c}{Propylene}\\
Energy range    & TDLDA (IPA) & Expt.      & TDLDA (IPA) & Expt.\\
\hline
10 eV $<\omega<25$ eV & 10.0 (11.1) & 9.9  & 9.5 (10.9) & 9.4\\
25 eV $<\omega<35$ eV & 2.9  (1.2) & 3.3  & 2.9 (1.8) & 3.1\\
35 eV $<\omega<60$ eV & 2.9  (2.3) &      & 2.7 (1.8) &    \\
\hline
\end{tabular}
\end{center}
\label{TRK}
\end{table*}
%%%%%%%%%%%%%%%%%%%%%%%%%%%%%%%%%%%%%%%%%%%%%%%%%%%%%%%%%%%%%%%%%%

\subsection{Origin of the difference in photoabsorption between the isomers}
We would like to discuss details of resonance peaks and an origin of
the different behaviors between the isomers.
First, let us compare the IPA results for the two isomers
(See dashed lines in Figs.~\ref{dfdw_cyclo1} (a) and \ref{dfdw_propy} (a)).
In the energy region of $10-20$ eV, although the bulk structure is
similar, cyclopropane shows a sharper main peak
at 11.5 eV and an additional peak at 13 eV.
This may be due to higher-fold degeneracies in electronic eigenstates
in cyclopropane.
These peak structures in cyclopropane remain after inclusion of the dynamical
screening effects, while, for propylene, the strong peak at 12 eV
seen in the IPA is diminished.
Next, we shall examine this difference in the dynamical screening effects.

For this purpose, it is useful to calculate a partial oscillator strength
\cite{ZS80,NY01}
which corresponds to a contribution of each occupied orbital to the total
oscillator strength.
We display the partial $df/d\omega$ in the energy range of $10-25$ eV
in Figs.~\ref{dfdw_cyclo1} (b) and \ref{dfdw_propy} (b).
One can see that the major contributions to $df/d\omega$
come from bound-to-continuum excitations of electrons near the Fermi level;
the HOMO $(3e')^4$ and the second HOMO $(1e'')^4$ in case of cyclopropane,
and the second $(10a')^2$, the third $(9a')^2$ and the fourth HOMO $(1a'')^2$
for propylene.

The sharp peaks at 11.8 eV and 13.5 eV in cyclopropane originate from
bound-to-bound transitions of $(3a_1')^2$ and $(1a_2'')^2$ electrons.
These resonances are also seen in the IPA calculation.
Then, the electron-electron correlation brings out coherent contributions
of bound-to-continuum excitations of $(3e')^4$ and $(1e'')^4$ electrons.
The width of the resonances becomes slightly larger than that of the IPA,
because of the autoionization process.
The peak at 15.7 eV is also produced by coherent excitations of
$(2e')^4$ (bound-to-bound) and $(3e')^4$ (bound-to-continuum) electrons.
We only see a shoulder around 15.5 eV in the IPA calculation, however,
the dynamical effect enhances the peak.

In the case of propylene,
we find several small peaks of bound-to-bound excitations in
the partial $df/d\omega$ in Fig.~\ref{dfdw_propy} (b).
However, the bound-to-continuum excitations, which mostly contribute to
the broad resonance in $13-18$ eV, behave rather independently,
and do not produce coherent enhancement of those peaks.
As a result, the small peaks in the bound-to-bound transitions are
mostly smeared out in the total oscillator strength.

We think that this difference in the continuum response could be
attributed to the difference in strength of bound-to-bound transitions.
The $df/d\omega$ of propylene shows a typical behavior of the dynamical
screening effects.
Namely, the oscillator strengths (and the peak at 12 eV) in the IPA calculation
are significantly weakened by the induced screening field in
Eq.~(\ref{delta_n}).
Conversely, those at energies above 16 eV are enhanced.
This is because the real part of the dynamical polarizability changes its
sign at $\omega\approx 16$ eV, then the screening field changes into
the ``anti-screening field'' at higher energies.
In the case of cyclopropane, the situation is slightly more complicated.
Generally speaking, the real part of the dynamical polarizability
changes its sign from positive to negative at bound resonances.
Since the degeneracies of electronic orbitals are higher in cyclopropane,
the bound-to-bound transitions have large oscillator strengths.
Then, there appears a reminiscence of bound resonance in the continuum
region.
The screening field suddenly drops down at energies corresponding to
bound-to-bound transitions.
This provides effective anti-screening effects to cause the peak structures
in the bound-to-continuum transitions.

\begin{figure}
\centerline{\includegraphics[width=0.4\textwidth]{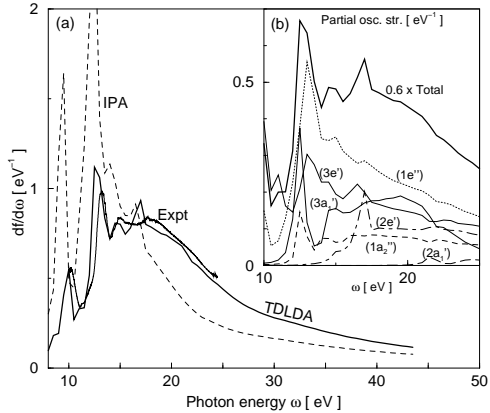}}
\caption{The same as Fig.~\ref{dfdw_cyclo1} but for using
a parameter $\beta=0.05$ in the calculation.
The calculation has been performed for an energy range of $8-43.5$
eV with a mesh of $\Delta\omega=0.5$ eV.
}
\label{dfdw_cyclo2}
\end{figure}

Finally, we would like to comment on dependence of our results upon the
parameter $\beta$ in the LB potential $\mu^{(\rm LB)}$.
A choice of this parameter is rather arbitrary, since the value of
$\beta$ does not change the Coulomb asymptotic behavior.
In fact, if we adopt $\beta=0.05$, the value proposed by the original
paper \cite{LB94},
we obtain better agreement to the photoabsorption spectra for both the isomers.
In the case of using $\beta=0.05$,
the Kohn-Sham eigenvalues for occupied orbitals in cyclopropane
are indicated as values in brackets in Table~\ref{spe}.
A binding energy of each orbital becomes deeper by $1.3-1.5$ eV,
though spacings between the orbitals almost stay constant.
The calculated oscillator strength distribution
is shown in Fig.~\ref{dfdw_cyclo2} for cyclopropane.
The disagreement on the peak positions are removed in the calculation.
A bound peak at 10 eV is also well reproduced in the calculation.
This peak consists of the excitations of $(3e')^4$ and $(1e'')^4$ electrons.
These excitations have an almost identical energy when we use $\beta=0.05$,
while the excitation of $(1e'')^4$ electrons is shifted to lower energy
by 1 eV when using $\beta=0.015$.
Except for the bound peak at 10 eV, the characteristic features of the
oscillator strength distribution are not changed, and we maintain
the interpretation given above.
%%%%%%%%%%%%%%%%%%%%%%%%%%%%%%%%%%%%%%%%%%%%%%%%%%%%%%%%%%%%%%%%%%
%%%%%%%%%%%%%%%%%%%%%%%%%%%%%%%%%%%%%%%%%%%%%%%%%%%%%%%%%%%%%%%%%%

\section{Conclusion}
\label{sec:conclusion}

The oscillator strength distributions of C$_3$H$_6$ isomer molecules
are studied with the TDLDA in the continuum utilizing the three-dimensional
Cartesian coordinate representation.
The calculation shows good agreement with experiments.
The oscillator strength in the energy region above 22 eV is almost identical
among the isomers, however, different peaks appear below that.
This isomer effect is analysed by calculating the partial oscillator strength
of each occupied orbital.
In addition to the difference in properties of bound electronic orbitals,
it turns out that bound-to-continuum excitations of electrons near the Fermi
level behave differently between the isomers.
The bound-to-bound transitions in cyclopropane possess large
strengths, and the bound-to-continuum transitions exhibit coherent peak
structures because of the anti-screening effects.
On the other hand, in propylene, the bound-to-bound transitions are too
weak to produce the anti-screening peaks for the continuum excitations.
Although the molecular structure directly has minor influence on the
continuum, the difference in bound-to-bound transitions
leads to variation in the dynamical screening effects to
change the continuum excitations.

\medskip

\noindent
{\bf Acknowledgements}

\medskip
This work is supported in part by Grants-in-Aid for Scientific Research
(No.1470146 and 14540369) from
the Japan Society for the Promotion of Science.
Calculations were performed on a NEC SX-5 Super Computer at Osaka University
and a HITACHI SR8000 at Institute of Solid State Physics, University of
Tokyo.

% \appendix

% \section{}
% \label{}


\begin{thebibliography}{00}

 \bibitem{Ber79}
J.~Berkowitz, Photoabsorption, Photoionization, and Photoelectron Spectroscopy,
Academic Press, New York, 1979.
 \bibitem{Hat99}
Y.~Hatano, Phys. Rep. 313 (1999) 109.
 \bibitem{Koi85}
H.~Koizumi, T.~Yoshimi, K.~Shinsaka, M.~Ukai, M.~Morita, Y.~Hatano,
J. Chem. Phys. 82 (1985) 4856.
 \bibitem{Kam99}
K.~Kameta, K.~Muramatsu, S.~Machida, N.~Kouchi, Y.~Hatano,
J. Phys. B 32 (1999) 2719.
\bibitem{LS84}
Z.~H.~Levine, P.~Soven, Phys. Rev. A 29 (1984) 625.
\bibitem{CCRM91}
I.~Cacelli, V.~Carravetta, A.~Rizzo, R.~Moccia,
Phys. Rep. 205 (1991) 283.
\bibitem{WL99}
M.~C.~Wells, R.~R.~Lucchese, J. Chem. Phys. 111 (1999) 6290.
\bibitem{Lan80}
P.~W.~Langoff, in: 
B.J. Dalton, S.M. Grimes, J.P. Vary, S.A. Williams (Eds.),
Theory and Application of Moment Methods in Many-Fermion Systems,
Plenum, New York, 1980, p. 191.
\bibitem{YM85}
S.~Yabushita, C.~W.~McCurdy, J. Chem. Phys. 83 (1985) 3547.
\bibitem{NY01}
T.~Nakatsukasa, K.~Yabana, J. Chem. Phys. 114 (2001) 2550.
\bibitem{ZS80}
A.~Zangwill, P.~Soven, Phys. Rev. A 21 (1980) 1561.
\bibitem{Ste89}
J.~J.~P.~Stewart, J. Comput. Chem. 10 (1989) 209.
\bibitem{PZ81}
J.~Perdew, A.~Zunger, Phys. Rev. B 23 (1981) 5048.
\bibitem{LB94}
R.~van~Leeuwen, E.~J.~Baerends, Phys. Rev. A 49 (1994) 2421.
\bibitem{Cas98}
M.~E.~Casida, C.~Jamorski, K.~C.~Caside, D.~R.~Salahub,
 J. Chem. Phys. 108 (1998) 4439.
\bibitem{TM91}
N.~Troullier, J.~L.~Martins, Phys. Rev. B 43 (1991) 1993.
\bibitem{KB82}
L.~Kreinman, D.~Bylander, Phys. Rev. Lett. 48 (1982) 1425.
\bibitem{Ple81}
V.~V.~Plemenkov, Y.~Y.~Villem, N.~V.~Villem, I.~G.~Bolesov, L.~S.~Surmina,
N.~I.~Yakushkina, A.~A.~Formanovskii, Zh. Obshch. Khim. 51 (1981) 2076.
\bibitem{Kra78}
D.~A.~Krause, J.~W.~Taylor, R.~F.~Fenske, J. Am. Chem. Soc. 100 (1978) 718.
\end{thebibliography}
\end{document}